\documentclass[preprint,showpacs,preprintnumbers,amsmath,amssymb]{revtex4}
\usepackage{graphicx}
\usepackage{dcolumn}
\usepackage{bm}

\begin{document}


\title{Dissociation of relativistic $^{10}$C nuclei in nuclear track
 emulsion}

\author{D.~A.~Artemenkov}
       \email{artemenkov@lhe.jinr.ru}  
   \affiliation{Joint Insitute for Nuclear Research, Dubna, Russia}
 \author{S.~S.~Alikulov}
   \affiliation{Joint Insitute for Nuclear Research, Dubna, Russia} 
\author{R.~R.~Kattabekov}
   \affiliation{Joint Insitute for Nuclear Research, Dubna, Russia} 
\author{K.~Z.~Mamatkulov}
   \affiliation{Joint Insitute for Nuclear Research, Dubna, Russia}  
 \author{N.~K.~Kornegrutsa}
   \affiliation{Joint Insitute for Nuclear Research, Dubna, Russia} 
\author{D.~O.~Krivenkov}
   \affiliation{Joint Insitute for Nuclear Research, Dubna, Russia} 
 \author{P.~I.~Zarubin}
     \email{zarubin@lhe.jinr.ru}    
     \homepage{http://becquerel.lhe.jinr.ru}
   \affiliation{Joint Insitute for Nuclear Research, Dubna, Russia}

\date{\today}

\begin{abstract}
Dissociation  of 1.2~A~GeV $^{10}$C nuclei in  nuclear track emulsionis is studied. 
 It is shown that most precise angular measurements provided by this technique play a crucial role
 in the restoration of the excitation spectrum of the 2$\alpha$+2p system. Strong contribution of the
 cascade process $^{10}$C$\rightarrow ^9$B$\rightarrow ^8$Be identified.\par
\end{abstract}

\maketitle
\section{\label{sec:level1}Introduction}
\indent The phenomenon of multiple fragmentation of relativistic nuclei can serve as a source of coherent
 ensembles of the lightest nuclei and nucleons. In this respect only nuclear track emulsion providing
 0.5~$\mu$m spacial resolution allow one to follow tracks of all relativistic fragments in forward cone defined 
 by a nucleon Fermi motion. The most peripheral collisions  accompanied by neither \lq\lq black\rq\rq nor
 \lq\lq gray\rq\rq tracks of target nucleus fragments are few percents  among inelastic interactions
 \cite{Andreeva05}. They are referred to as the \lq\lq white\rq\rq stars what also aptly reflects the short drop
 of ionization from  primary to secondary tracks. Such events occur  as a result of  electromagnetic and nuclear
 diffraction on heavy nuclei of emulsion composition (i.e., Ag and Br). Minimal  perturbation of a projectile
 make them the most valuable sample for nuclear cluster physics. Excitation energy of a fragment ensemble is
 estimated as Q = M$^*$-M, where M$^*$ is the ensemble invariant mass  and M - a projectile mass.
 The value M$^*$ is defined as M$^{*2}$ = ($\sum$P$_j$)$^2$=$\sum$(P$_i \cdot$P$_k$), where P$_{i,k}$ 
 are 4-momenta of the fragments. Assumption of projectile speed conservation by relativistic fragments is sufficient
 to compensate the lack of momentum measurements. Already it is established that final states of relativistic
 He fragments  effectively correlate with the clustering in the nuclei $^{12}$C \cite{Belaga95},
 $^{6}$Li \cite{Adamovich99}, and $^{9}$Be \cite{Artemenkov07}, \cite{ArtemenkovAIP07}, \cite{Artemenkov08}. The described approach is used in the BECQUEREL Project \cite{Web2010} to study  the dripline nuclei $^7$Be \cite{ArtemenkovAIP07,Peresadko07}, 
 $^{8}$B \cite{ArtemenkovAIP07,Stanoeva08}, $^{9}$C \cite{Krivenkov10}, $^{10}$C \cite{Kattabekov10}, and $^{12}$N \cite{Kattabekov10} by means emulsion stacks exposed to secondary beams of the JINR Nuclotron \cite{Rukoyatkin08}. The  status of the $^{10}$C investigation, which entails the production of two $\alpha$ particles and two protons, is presented.\par
 \par 
  \begin{figure}
  \includegraphics[width=0.95\textwidth]{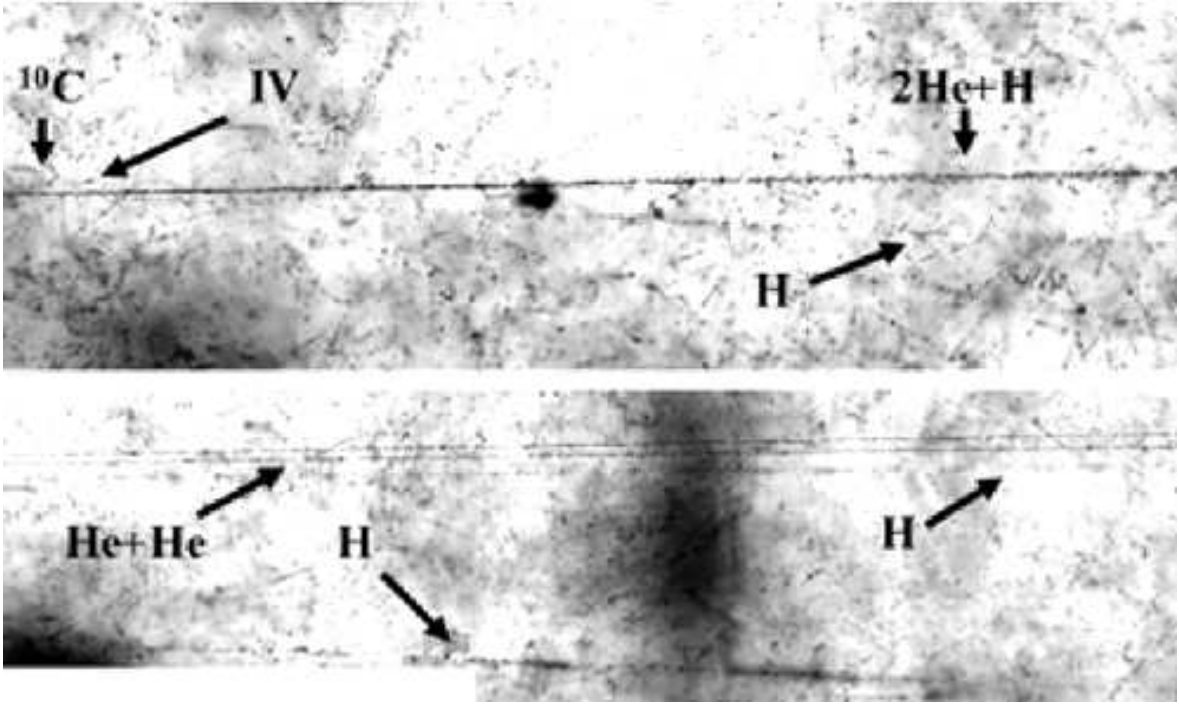}
  \caption{Dissociation $^{10}$C$\rightarrow$2He+2H  
 in a nuclear track emulsion (\lq\lq white\rq\rq ~star).
  The interaction vertex is indicated as IV and secondary tracks as  H and He.}
\label{fig:1}       
 \end{figure} 

\section{\label{sec:level2}Experiment}
\label{sec:1}
\indent Nuclear track emulsion is exposed to a mixed beam of  $^{12}$N, $^{10}$C and $^7$Be nuclei 
 formed by means of  primary 1.2~A~GeV  $^{12}$C nucleus beam \cite{Kattabekov10},\cite{Rukoyatkin08}.  The scanning along the
 total length of primary tracks in emulsion layers that was equal to 924.7 m revealed 6144 inelastic interactions, including 330 
\lq\lq white\rq\rq stars.
 For \lq\lq white\rq\rq stars with charge topology $\sum$Z$_{fr}$=6 the most probable channel is
 represented by 159 events 2He+2H as expected for the $^{10}$C isotope. Example of such event is shown in Fig.~\ref{fig:1}. The channel He+4H is found to
 be suppressed (16 events) since it has  higher threshold for an
 $\alpha$-cluster break up.\par
 
\begin{figure*}
  \includegraphics[width=0.5\textwidth]{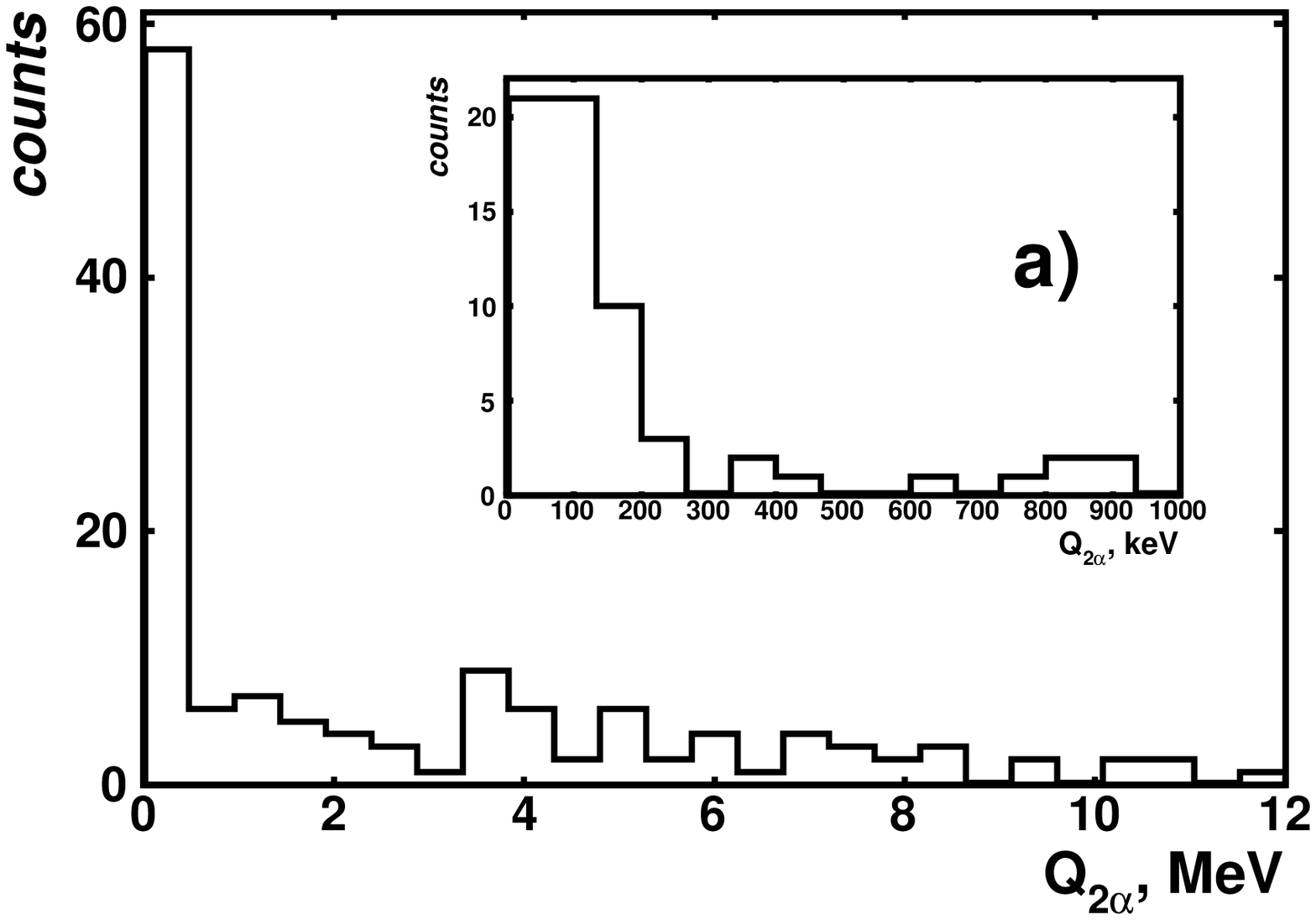}
  \includegraphics[width=0.5\textwidth]{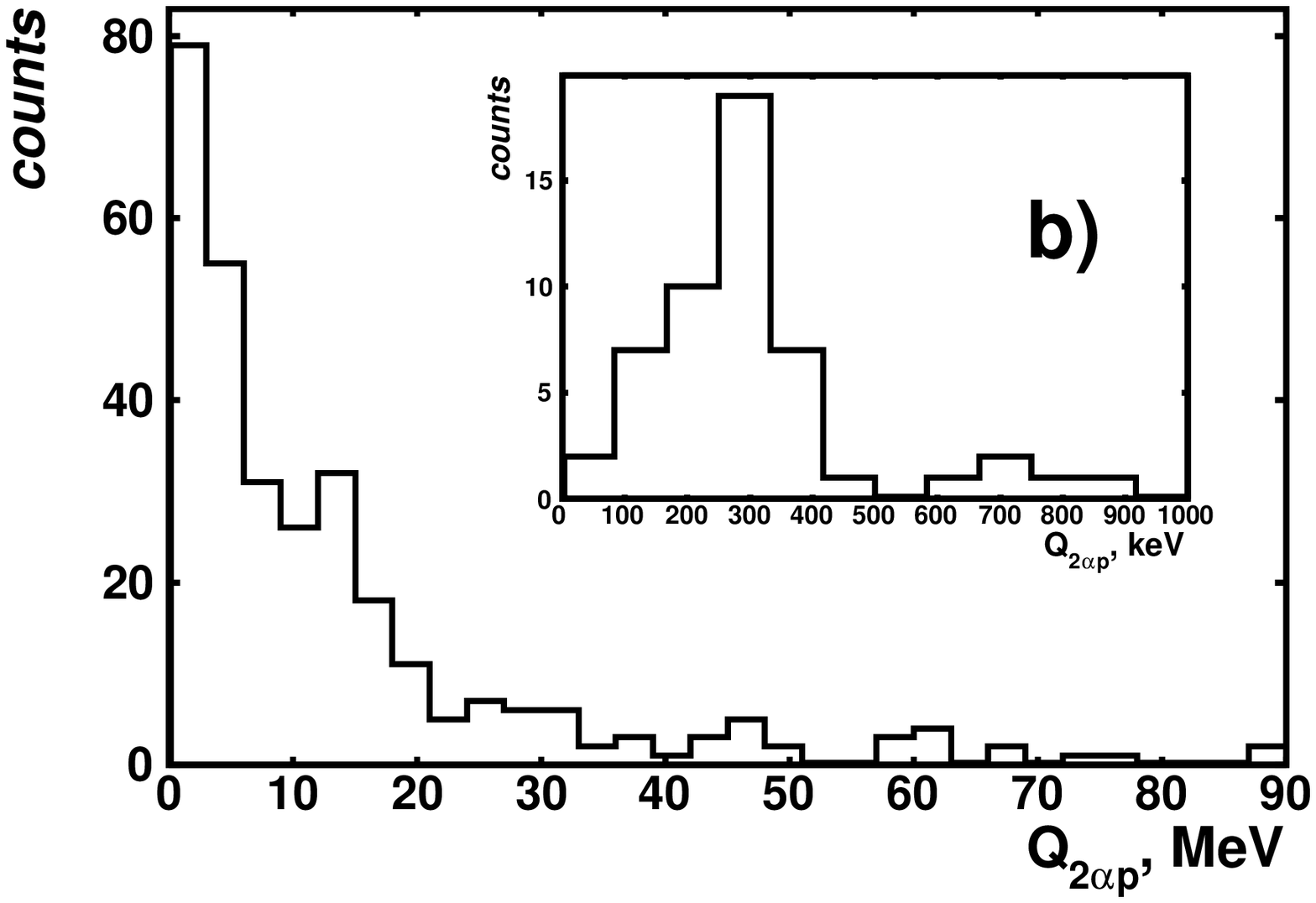}
\caption{ a)Distribution of the number of \lq\lq white\rq\rq stars  2$\alpha$+2p 
 versus  excitation energy Q$_{2\alpha}$ of the $\alpha$-pairs. In the inset a zoom over the Q$_{2\alpha}$ distribution is shown.
 b) Distribution of the number of \lq\lq white\rq\rq stars 2$\alpha$ + 2p 
 versus  excitation energy Q$_{2\alpha p}$ of triples 2$\alpha$ + p. In the inset a zoom over the Q$_{2\alpha p}$ distribution is shown.}
\label{fig:2}       
\end{figure*}

\indent  The core role of the unbound $^8$Be nucleus in the $^{10}$C structure is
 manifested in intensive fragmentation via $^{10}$C$\rightarrow ^8$Be+2p.  Distribution of $\alpha$-pairs in the 156
 \lq\lq white\rq\rq stars 2$\alpha$ + 2p on the excitation energy Q$_{2\alpha}$ is presented in Fig.~\ref{fig:2}~a.
 In 63 events the Q$_{2\alpha}$ value does not exceed 1 MeV (inset in Fig.~\ref{fig:2}~a ). For them, the average value is
 $<Q_{2\alpha}>\approx$63$\pm$30~keV and the mean-square scattering $\sigma$=83~keV, which agrees
well with the decay of the $^8$Be 0$^+$ ground state. The $^8$Be 0$^+$ contribution is approximately the same as for the 
 neighboring cluster nuclei \cite{Belaga95}, \cite{ArtemenkovAIP07}.\par
\indent The unbound $^9$B nucleus can be another major product of the $^{10}$C coherent dissociation. Fig.~\ref{fig:2}~b shows
the distribution of \lq\lq white\rq\rq stars 2$\alpha$+2p on the excitation energy Q$_{2\alpha p}$, defined by the
 difference of the invariant mass of the three fragments 2$\alpha$+p and the mass of the proton and the doubled 
 $\alpha$-particle masses. The Q$_{2\alpha p}$ values for one of two possible 2$\alpha$+p triples do not exceed
 1~MeV in 58 events (inset in Fig.~\ref{fig:2}~b). The average value for these triples is $<Q_{2\alpha p}>$=254$\pm$18~keV
 with rms $\sigma$=96~keV. These values correspond well to the $^9$B ground state decay via the channel
 p+$^8$Be (0$^+$) with energy 185~keV and width (0.54$\pm$0.21)~keV. In the
 region limited by Q$_{2\alpha}<$1~MeV and Q$_{2\alpha p}<$1~MeV there is a clear correlation in the $^8$Be and
 $^9$B production. One can note the formation of a single event 2$\alpha$+2p with Q$_{2\alpha p}$ equal to the
 values 0.23 and 0.15~MeV, i.e., at the same time both triples correspond to the decay of the nucleus $^9$B (Fig.~\ref{fig:2}~b). In
 all other $^9$B cases one of  Q$_{2\alpha p}$ is above 1 MeV. Excitation channel $\alpha$+2p is studied on the remaining statistics of \lq\lq white\rq\rq stars
 2$\alpha$+2p beyond $^9$B decays. There is no clear signal of $^6$Be decays.\par

\section{\label{sec:level3}Conclusion}

\indent To conclude,
 contribution of $^8$Be nuclei is about one-third in ralativistic $^{10}$C dissociation.
The production of $^8$Be nuclei shows  strong correlation 
with cascade decay $^{10}$C $\rightarrow ^9$B$\rightarrow ^8$Be.
 There is no significant contribution of decays $^8$Be$\rightarrow$2$\alpha$ through the first excited
 state 2$^+$, which differs qualitatively the $^{10}$C and $^9$Be nuclei
\cite{Artemenkov07}, \cite{ArtemenkovAIP07}, \cite{Artemenkov08}. It can be assumed that the
 $^8$Be 2$^+$ state does not contribute to the ground state of the $^{10}$C nucleus, and its core is formed of
 the 0$^+$ state. Paired protons can have the meaning of a covalent pair in the $^{10}$C molecular-like
 system with two-center potential $\alpha$+2p+$\alpha$. Verification of these assumptions will be made in the
 correlation analysis of the pairs of 2p and 2$\alpha$ is foreseen with the rest two-thirds of statistics.\par


\end{document}